\documentclass[prl,twocolumn,superscriptaddress,showpacs,floatfix,longbibliography,aps,10pt]{revtex4-2}

\newif\ifarxiv\arxivtrue

\def\minisection#1{\ifarxiv\section*{#1}\else\par\vskip20pt\noindent{\bf #1}\ \fi}
\def\minisubsection#1{\ifarxiv\subsection*{#1}\else\par\vskip20pt\noindent{\bf #1}\ \fi\noindent}

\usepackage{mathrsfs,braket}
\usepackage{amssymb, amsbsy, amsmath, latexsym, dsfont, array, layout,
graphicx,mathrsfs,xcolor, bm}
\usepackage{cancel}
\usepackage[colorlinks=true,citecolor=blue,urlcolor=blue]{hyperref}

\newcommand{\ketbra}[2]{\left|{#1}\rangle\!\langle{#2}\right|}

\usepackage{changes}
\DeclareMathOperator{\arcsinh}{arcsinh}

\begin{document}

\title{Observation of Metal-Insulator and Spectral Phase Transitions in Aubry-Andr\'{e}-Harper Models}

\author{Quan Lin} \thanks{These authors contributed equally to this work.}
\affiliation{School of Physics, Southeast University, Nanjing 211189, China}

\author{Christopher Cedzich} \thanks{These authors contributed equally to this work.}
\affiliation{Heinrich Heine Universit\"at D\"usseldorf, Universit\"atsstr. 1, 40225 D\"usseldorf, Germany}
\affiliation{Fakult\"at f\"ur Mathematik und Informatik, FernUniversit\"at in Hagen, Universit\"atsstr. 1, 58097 Hagen, Germany}

\author{Qi Zhou}
\affiliation{
Chern Institute of Mathematics and LPMC, Nankai University, Tianjin 300071, China
}

\author{Peng Xue}\email{gnep.eux@gmail.com}
\affiliation{Beijing Computational Science Research Center, Beijing 100193, China}

\begin{abstract}
{\bf Non-Hermitian extensions of the Aubry-Andr\'{e}-Harper (AAH) model reveal a rich variety of phase transitions arising from the interplay of quasiperiodicity and non-Hermiticity. Despite their theoretical significance, experimental explorations remain challenging due to complexities in realizing controlled non-Hermiticity. Here, we present the first experimental realization of the unitary almost-Mathieu operator (UAMO) which simulates the AAH model by employing single-photon quantum walks. Through precise control of quasiperiodicity, we systematically explore the phase diagram displaying a phase transition between localized and delocalized regimes in the Hermitian limit. Subsequently, by introducing non-reciprocal hopping, we experimentally probe the parity-time (PT) symmetry-breaking transition that is characterized by the emergence of complex quasienergies. Moreover, we identify a novel spectral transition exclusive to discrete-time settings, where all quasienergies become purely imaginary. Both transitions are connected to changes in the spectral winding number, demonstrating their topological origins. These results clarify the interplay between localization, symmetry breaking, and topology in non-Hermitian quasicrystals, paving the way for future exploration of synthetic quantum matter.}
\end{abstract}

\maketitle

Phase transitions are ubiquitous in nature and manifest themselves in various forms, each characterized by distinct physical signatures. Their defining feature is the abrupt change in an observable quantity such as an order parameter or characteristic length scale, which is sometimes accompanied by a spontaneous breaking of symmetries within the system. Notable examples include transitions between thermodynamic states of matter~\cite{stanley1971phase}, phase transitions in topological insulators and superconductors~\cite{hasanColloquiumTopologicalInsulators2010,kitaevPeriodicTableTopological2009,qiTopologicalInsulatorsSuperconductors2011,ryuTopologicalInsulatorsSuperconductors2010}, and disorder-induced metal-insulator transitions in low-dimensional random systems~\cite{ander}. The latter describes the phenomenon that for disorder stronger than a certain ``critical'' value, wave functions become localized, impeding their diffusion across the medium. This so-called ``Anderson localization'' has been extensively studied via theoretical and experimental approaches across diverse media~\cite{FiftyYearsOfAL,kunzSpectreOperateursAux1980,stolzIntroductionMathematicsAnderson2011,LeeRamakrishnan1985,sokoloff,roati,Segev,lahini,lin1,lin2}.

Among the numerous models elucidating such a localization-delocalization transition, the Aubry-Andr\'{e}-Harper (AAH) model stands out due to its intrinsic Andr\'{e}-Aubry duality and its emblematic role in studying quasicrystals~\cite{lahini,Segev,roati,AA,fliu,sahu,Jitomirskaya1999Annals,sokoloff,verbin,szame,Lahini2007}.
Characterized by an incommensurate potential applied to one-dimensional lattices, the AAH model exhibits sharp metal-insulator transitions driven by variations in the potential strength~\cite{AA,Jitomirskaya1999Annals,avilaSharpPhaseTransitions2017}.
These investigations have provided key insights into the model's utility and its links to other pivotal frameworks, such as the Hofstadter model~\cite{brownBlochElectronsUniform1964,hofstadter} which describes the motion of electrons in a homogeneous magnetic field on a two-dimensional lattice and plays an important role in the understanding of quantum Hall systems~\cite{thoulessQuantizedHallConductance1982,niuQuantizedHallConductance1985}. The universality of the AAH model is further highlighted by the broad experimental observation of metal-insulator transitions across multiple physical platforms, underscoring their significance in contemporary condensed-matter research~\cite{lahini,Segev,roati}.

\begin{figure*}[th]
\includegraphics[width=0.95\textwidth]{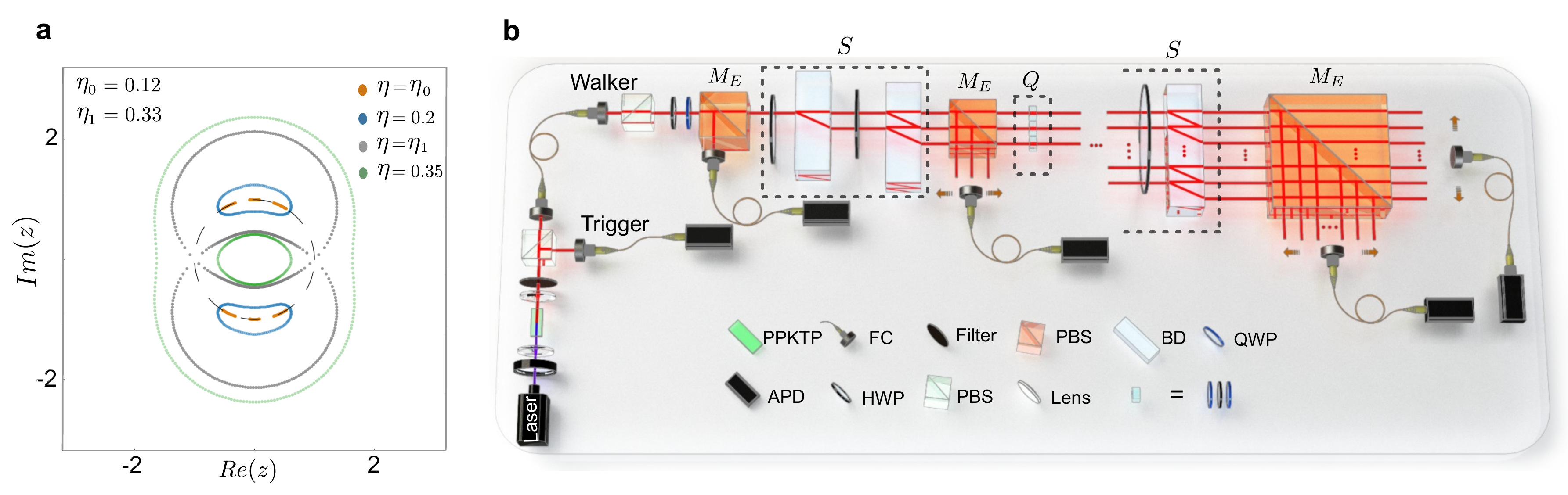}
\caption{{\bf Spectra of Floquet operator $W$ and experimental implementation.} {\bf a} The eigenvalue $z$ of $W$ for $(\lambda_{1}, \lambda_{2})=(0.25, 0.5)$, with different colors corresponding to variations in $\eta$. For visual guidance, a unit circle (dashed black line) is included. {\bf b} A pair of photons is generated through spontaneous parametric down-conversion in a periodically poled potassium titanyl phosphate (PPKTP) crystal, with one photon serving as a trigger and the other as the walker in the quantum-walk network. The initial state is prepared using a polarizing beam splitter (PBS), a quarter-wave plate (QWP), and a half-wave plate (HWP). It then undergoes a quantum walk through an interferometric network comprising HWPs, beam displacers (BDs), and partially polarizing beam splitters (PPBSs). It is eventually detected by avalanche photodiodes (APDs), in coincidence with the trigger photon. }\label{fig1}
\end{figure*}

Recent theoretical advancements have significantly broadened the understanding of phase transitions within the AAH model and its non-Hermitian extensions~\cite{xcai,longhi21,yliu,yucePTSymmetricAubry2014,longhi19,jiangInterplayNonHermitianSkin2019}. An essential property of these extensions is the presence of parity-time (PT) symmetry, along with a phase transition known as the spontaneous breaking of PT symmetry, up to which the spectrum remains real even though the system remains non-Hermitian~\cite{benderIntroductionPTSymmetricQuantum2005,benderRealSpectraNonHermitian1998,footnote1}.
Despite these theoretical advances, experimental realizations of the AAH model and its non-Hermitian extensions in optical systems have remained inaccessible, largely due to challenges in engineering quasiperiodic drives with controlled non-Hermiticity. To overcome these challenges, we investigate the unitary almost-Mathieu operator (UAMO)~\cite{cedzi}. The UAMO has been rigorously demonstrated to be an exactly solvable simulator of the AAH model, exhibiting analogous metal-insulator phase transitions and localization properties~\cite{cedzi,twentyMartinis,cedzich2024absence}. We further extend this framework to the non-Hermitian regime by introducing a gain-loss parameter into the UAMO~\cite{cedzi2}. The exact solvability of the UAMO facilitates precise theoretical predictions and experimental verifications, further substantiating its potential for wide-ranging applications in the exploration of quantum systems.

In this work, we report the first experimental implementation of the UAMO and its non-Hermitian extension using dynamic signatures in single-photon quantum walks~\cite{xiao1,xiao2}.
We observe metal-insulator transitions driven by variations in the coupling constants. Moreover, we confirm the presence of a novel spectral transition that was theoretically predicted in~\cite{cedzi2}, characterized by all quasienergies acquiring imaginary components. We correlate the non-Hermitian phase transitions with features of the eigenvalue spectrum, which can be characterized by the change of a topological winding number emerging from the closed contours of the spectrum in the complex plane. Our results provide the first experimental demonstration of the non-Hermitian AAH model and reveal the rich interplay between quasiperiodicity, non-Hermiticity, and discrete-time dynamics. Our platform not only facilitates precise control over phase transitions but also paves the way for designing synthetic structures with customized transport properties and for exploring novel topological phases in non-Hermitian systems.

\begin{figure*}[th]
\includegraphics[width=\textwidth]{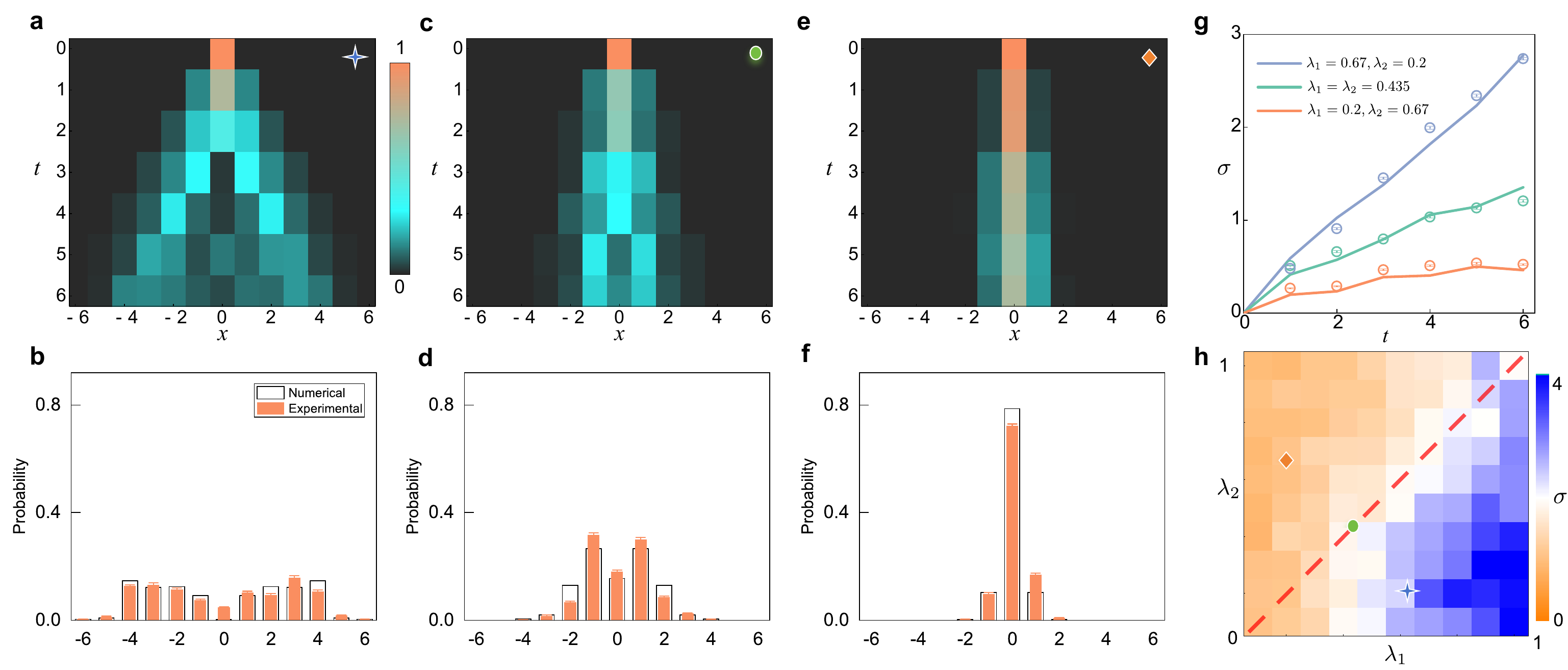}
\caption{{\bf Experimental observation of metal-insulator transition in the AAH model.} {\bf a}, {\bf c}, {\bf e} Measured probability distribution $P(x, t)$ of a 6-step Hermitian quantum walk governed by $W$ with $\lambda_1=0.67>\lambda_2=0.2$, $\lambda_1=\lambda_2=0.435$, $\lambda_1=0.2<\lambda_2=0.67$, respectively. {\bf b}, {\bf d}, {\bf f} Measured and simulated probability distributions at the last step, $P(x, t =6)$, corresponding to {\bf a}, {\bf c}, {\bf e}, respectively. {\bf g} Measured (symbols) and simulated (lines) standard deviation $\sigma$ of the Hermitian quantum walk. {\bf h} The measured phase diagram evaluated by the standard deviation $\sigma$ at time step $t = 6$. Note that for small values of $\lambda_1$, the walker hardly moves, and for short times such as $t=6$ is virtually indistinguishable from a localized walker. Hence, the phase boundary in the bottom-left region is not sharply defined. For all plots, we set $\theta=0$ and used the initial state $\ket{\psi(0)}=\ket{0}\otimes(\ket{H}+i\ket{V})/\sqrt{2}$. Error bars are due to the statistical uncertainty in photon number counting.}\label{fig2}
\end{figure*}

\minisection{Results}
\minisubsection{The pseudo-unitary almost Mathieu operator}

We investigate a non-unitary quantum walk on a one-dimensional lattice with two-dimensional spin at each lattice site using single photons within an alternating-loss scheme.
The dynamics of the system are obtained by iteratively applying the Floquet operator
\begin{align}
W_{\lambda_{1},\lambda_{2},\theta,\eta}=S_{\lambda_{1},\eta}Q_{\lambda_{2},\theta}
\label{eq:U}
\end{align}
to an initial state $|\psi(0)\rangle$, resulting in the time-evolved state $|\psi(t)\rangle=W_{\lambda_{1},\lambda_{2},\eta}^t|\psi(0)\rangle$, where $t$ labels the discrete time steps.
Here $S_{\lambda_{1},\eta}$ is the shift operator, i.e., $S_{\lambda_{1},\eta}=\sum_{x}( e^{2\pi \eta} \lambda_{1}|x+1\rangle\langle x| \otimes|0\rangle\langle 0|-\lambda_{1}^{\prime}|x\rangle\langle x| \otimes|0\rangle\langle 1|+\lambda_{1}^{\prime}|x\rangle\langle x| \otimes|1\rangle\langle 0|+ e^{-2\pi \eta} \lambda_{1}|x-1\rangle\langle x| \otimes|1\rangle\langle 1|)$, where $0\leq\lambda_1\leq1$ is called the {\it coupling constant}, as it controls the strength of coupling between neighboring lattice sites, and $\lambda_{1}^{\prime}=\sqrt{1-\lambda_{1}^{2}}$.
The parameter $\eta$ quantifies the imbalance of $S_{\lambda_1,\eta}$ between left-moving and right-moving modes. For $\eta=0$, the shift is balanced and, in particular, unitary.
Whenever $\eta\neq0$, the shift is not unitary anymore in the standard sense. By convention, we refer to these regimes as the ``Hermitian'' and the ``non-Hermitian'' regime, respectively, and we call $\eta$ the ``non-Hermitian parameter''. This nomenclature is adopted to align with the literature~\cite{lin2,szame}, where a non-vanishing $\eta$ breaks Hermiticity. 

The coin operator $ Q_{\lambda_{2},\theta}=\sum_{x}|x\rangle\langle x|\otimes Q_x $ acts locally via a quasiperiodic matrix $Q_x\in\mathrm{SU}(2)$ given by
\begin{align}
Q_{x}= \begin{bmatrix}\lambda_2\cos(2\pi(x\Phi+\theta))+i\lambda_2^{\prime}&-\lambda_2\sin(2\pi(x\Phi+\theta))\\\lambda_2\sin(2\pi(x\Phi+\theta))&\lambda_2\cos(2\pi(x\Phi+\theta))-i\lambda_2^{\prime}\end{bmatrix}
\label{eq:Q}
\end{align}
with $\lambda_2^{\prime}=\sqrt{1-\lambda_2^{2}}$. Here, $\Phi \in [0,1]$ plays the role of a magnetic field in an associated two-dimensional system~\cite{cedzi3}, $0\leq\lambda_2\leq1$ controls the coupling of the shift in the synthetic dimension, and the phase $ \theta \in [0,1]$ is its Fourier parameter~\cite{cedzi}. In what follows, the dependence on $\theta$ is omitted in the notation.

In the Hermitian setting $\eta=0$, the model is called UAMO~\cite{cedzi} and has an Andr\'{e}-Aubry duality with dual $W_{\lambda_1,\lambda_2,0}^\sharp=W_{\lambda_2,\lambda_1,0}^\top$. The arithmetic properties of $\Phi$ play a crucial role: for rational values $\Phi = n/m$, the system exhibits periodic behavior, and the evolved states propagate ballistically, with the spectrum of $W$ consisting of $2m$ bands that resemble a twofold copy of the Hofstadter butterfly. In this work, we focus exclusively on irrational fields, specifically $\Phi=(\sqrt{5}-1)/2$. Under these conditions, the UAMO $W_{\lambda_1,\lambda_2,0}$ possesses a metal-insulator phase transition about the self-dual line $\lambda_1=\lambda_2$. More specifically, for $\lambda_1>\lambda_2$, one observes ballistic transport, whereas, in the Aubry-dual regime $\lambda_1<\lambda_2$, the system displays Anderson localization~\cite{cedzi,cedzi1}.
This can be inferred from the Lyapunov exponent $L_{\lambda_1,\lambda_2,\Phi}$ of the model, which describes the average decay of eigenstates (see Supplemental Material). On the spectrum, it vanishes for $\lambda_1\geq\lambda_2$.
In the localized phase, we have $L_{\lambda_1,\lambda_2,\Phi}=\log\lambda_0>0$, where
\begin{equation}\label{eq:lambda0}
    \lambda_0=\frac{\lambda_2(1+\lambda_1')}{\lambda_1(1+\lambda_2')}.
\end{equation}
Given this phenomenological equivalence between the UAMO and the AAH model, we use the quantum walk $W_{\lambda_1,\lambda_2,0}$ to probe the properties of the AAH model.

\begin{figure*}[th]
\includegraphics[width=0.75\textwidth]{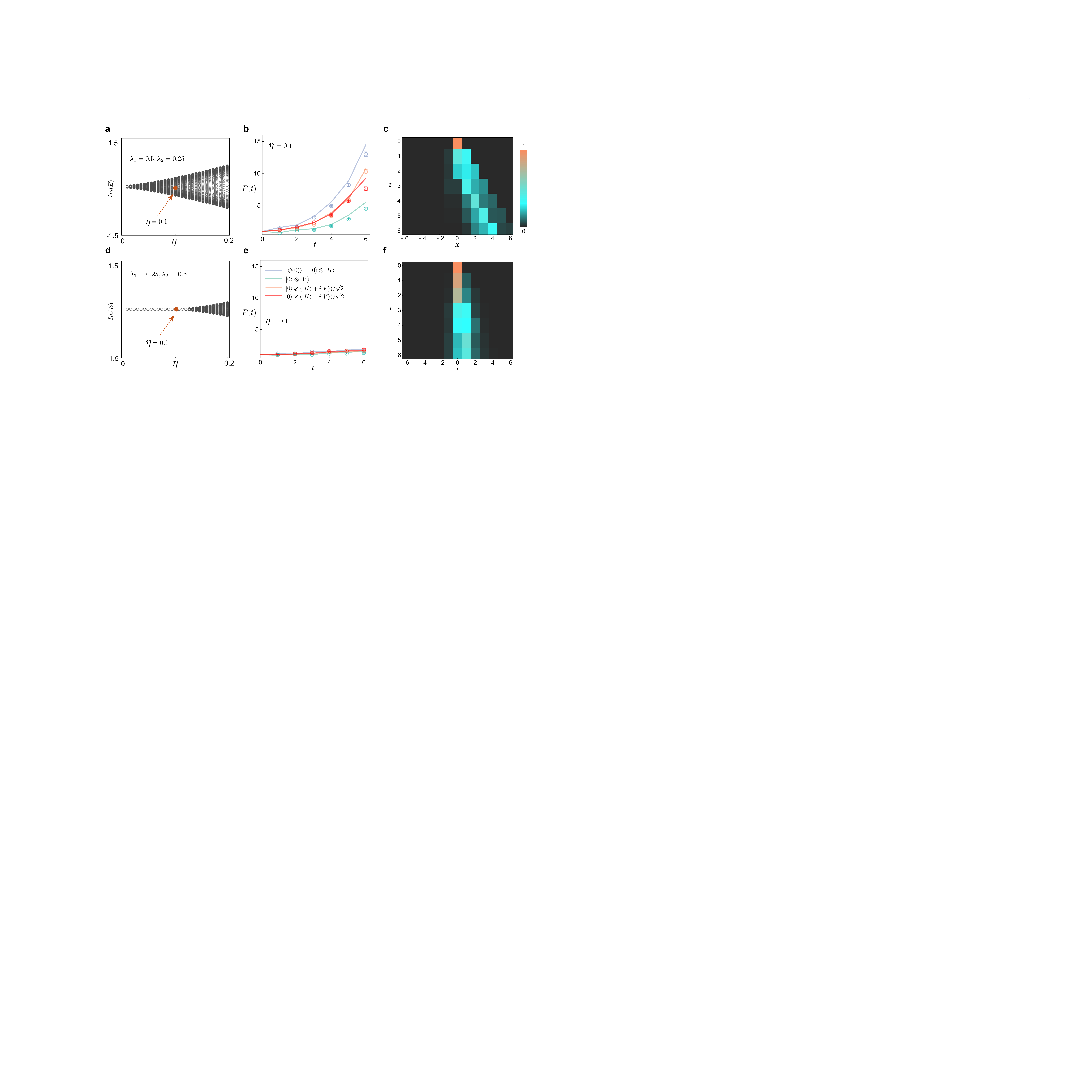}
\caption{{\bf Observation of the PT-symmetry transition in the non-Hermitian AAH model.} {\bf a}, {\bf d} Imaginary parts of the quasienergies for the non-Hermitian quantum walk with $\theta=0$ for $\lambda_1=0.5>\lambda_2=0.25$ and $\lambda_1=0.25<\lambda_2=0.5$, respectively. {\bf b}, {\bf e} Measured overall probabilities $P(t)$ of the quantum walk with different initial states corresponding to the PT unbroken and broken phases, respectively. The parameters in {\bf b}, {\bf e} are the same as those in {\bf a} and {\bf d}, respectively. {\bf c}, {\bf f} Measured probability distributions $P(x, t)$ of the 6-step non-Hermitian quantum walk corresponding to {\bf b} and {\bf e}, the initial state for each is $\ket{\psi(0)}=\ket{0}\otimes(\ket{H}+i\ket{V})/\sqrt{2}$.}\label{fig3}
\end{figure*}

In the non-Hermitian extension of the AAH model with unbalanced hopping, a topological phase transition between the localized and the delocalized phase occurs once the PT symmetry is broken~\cite{longhi19,yliu}. This extension is realized in our system by introducing the non-Hermitian parameter $\eta\neq0$, and $W_{\lambda_1,\lambda_2,\eta\neq0}$ is called the {\it pseudo}-UAMO (PUAMO)~\cite{cedzi2,footnote2}. We find that the PUAMO is PT symmetric (see Methods) and possesses a PT symmetry-breaking phase transition that is signaled by some quasienergies of the PUAMO acquiring imaginary components. This phase transition is topological and can be characterized by the winding number~\cite{cedzi2}
\begin{align}
\nu_\eta(z)=\lim_{N\to\infty}\frac{1}{N}\frac{1}{2\pi i}\int_0^1d\theta\partial_\theta\log\det(W_{\lambda_1,\lambda_2,\eta}-z),
\label{eq:w}
\end{align}
where $N$ is the lattice size, and $z$ with $|z|=1$ is a base point in a gap of the spectrum of the UAMO~\cite{twentyMartinis}. The winding number $\nu_\eta(z)$ is quantized and can take only three values. When $\nu_\eta(z) = 0$, the PUAMO retains unbroken PT symmetry for $\log \lambda_0 < \eta < -\log \lambda_0$ (see Supplemental Material). As illustrated in Fig.~\ref{fig1}{\bf a}, by increasing the non-Hermitian parameter $\eta$ beyond the critical point $\eta_{\text{PT}}=-\log\lambda_0$, the PT symmetry is spontaneously broken, and a condition emerges for $z$ in the unit circle such that $|\nu_\eta(z)|=1$. This phase transition is analogous to the spontaneous breaking of PT symmetry in the non-Hermitian AAH model. Further increasing $\eta$ triggers a novel spectral phase transition at $\eta_0=\arcsinh(\lambda_1'/\lambda_1)/(2\pi)$, characterized by purely imaginary quasienergies and $|\nu_\eta(z)|=1$ for all $z$ on the unit circle~\cite{cedzi2}. This spectral transition does not appear in the pervious non-Hermitian AAH model~\cite{cedzi2,footnote2} (see Supplemental Material). In combination with the metal-insulator transition in the Hermitian case, we systematically explore these transitions experimentally, as detailed in the subsequent sections.

\minisubsection{Experimental detection of metal-insulator phase transitions in the UAMO model}

We implement the PUAMO with single photons using the photonic setup illustrated in Fig.~\ref{fig1}{\bf b}. The coin states $|0\rangle$ and $|1\rangle$ are encoded in the horizontal $|H\rangle$ and vertical $|V\rangle$ polarizations of photons, respectively, while their spatial modes represent the lattice degrees of freedom. The coin-rotation operators $Q_{\lambda_2,\theta}$ and the shift operator $S_{\lambda_1,\eta}$ are implemented using a combination of HWPs, BDs, and PPBSs. Photon detection is performed by APDs, allowing us to resolve the walker's spatial distribution at each time step (see Methods).

We first focus on the Hermitian regime ($\eta=0$), where theoretical studies in~\cite{cedzi} predict a metal-insulator transition that has not yet been explored experimentally. As illustrated in Figs.~\ref{fig2}{\bf a}, {\bf b}, we experimentally observe a ballistic distribution in the quantum-walk dynamics, where the wave packet spreads linearly in time, confirming that the system is in the metallic phase for $\lambda_1 > \lambda_2$. Conversely, when $\lambda_1<\lambda_2$, we observe that the system is insulating, and wave packets remain localized (see Figs.\ref{fig2}{\bf e}, {\bf f}). These two distinct dynamic behaviors are demarcated by the dynamics at the critical region $\lambda_1=\lambda_2$, as shown in Figs.~\ref{fig2}{\bf c}, {\bf d}. We compare the measured photostatistics with the theoretical prediction via the {\it similarity} $\mathcal{S}(t)=\big[\sum_x\sqrt{P_\text{exp}(x,t)P_\text{th}(x,t)}\big]^2$, ranging from $0$ to $1$ for completely orthogonal and identical distributions, respectively. Here $P(x,t)=\big(|\langle{x}|\otimes\langle{0}|\psi(t)\rangle|^2+|\langle{x}|\otimes\langle{1}|\psi(t)\rangle|^2\big)/\sum_x\big(|\langle{x}|\otimes\langle{0}|\psi(t)\rangle|^2+|\langle{x}|\otimes\bra{1}\psi(t)\rangle|^2\big)$ is the probability at lattice site $x$ and time $t$, and the subscripts ``exp'' and ``th'' denote experimentally measured and theoretically predicted probability distributions, respectively. In our experiments, we achieve $\mathcal{S}>0.961$ in Fig.~\ref{fig2}{\bf a}, $\mathcal{S}>0.978$ in Fig.~\ref{fig2}{\bf c}, and $\mathcal{S} >0.984$ in Fig.~\ref{fig2}{\bf e} for all $t=1,\dots,6$.

The localization transition can also be quantified by assessing the standard deviation of dynamic probabilities, defined as $\sigma(t) = \sqrt{\sum_{x}P(x,t)x^{2}-(\sum_{x}P(x,t)x)^2} $. In Fig.~\ref{fig2}{\bf g}, we observe that $\sigma$ exhibits linear growth ($\sigma \sim t$) in the metallic phase with $\lambda_{1}=0.67>\lambda_2=0.2$, whereas $\sigma$ remains constant ($\sigma \sim \mathrm{const}.$) in the insulating phase with $\lambda_1 = 0.2 < \lambda_2 = 0.67$. These observations align well with theoretical predictions, as illustrated by the lines in Fig.~\ref{fig2}{\bf g}. To gain further insight of the phase transition, we experimentally map the relationship between the standard deviation $\sigma$ and the parameters $\lambda_{1}, \lambda_{2}$ by scanning the whole parameter space. The distinct characteristics of $\sigma$ in different phase regions enables clear identification of the phase regions after quantum-walk steps $t=6$. Our experimental results, as shown in Fig.~\ref{fig2}{\bf h}, graphically depict a phase diagram that explicitly delineates the regions corresponding to the two distinct dynamical behaviors. These delocalized and localized regions are clearly separated by the symmetric coupling $\lambda_1 = \lambda_2$, as shown in Fig.~\ref{fig2}{\bf h}. Along this self-dual line, all eigenstates of $W_{\lambda_1,\lambda_2}$ are critical, exhibiting neither delocalized nor localized behavior, and are characterized by a singular continuous quasienergy spectrum~\cite{cedzi}.

\minisubsection{Observation of the spontaneous breaking of PT symmetry in the PUAMO model}

We now turn to the non-Hermitian extension of the UAMO by introducing a non-zero non-Hermitian parameter $\eta$ to simulate unbalanced hopping in the AAH model.
Notably, through generalized Aubry duality, this is equivalent to introducing a complex quasiperiodic phase in $Q_{\lambda_1,\theta}$, i.e., $\theta \mapsto \theta-i\eta$ (see~\cite{cedzi2} for details). This correspondence thus gives experimental access to complexified quasiperiodic phases.

When non-Hermiticity is introduced in the metallic phase ($\lambda_{1}>\lambda_{2}$), the PT symmetry of PUAMO is broken for any non-zero $\eta$, as demonstrated in Fig.~\ref{fig3}{\bf a}. Here, the quasienergy $E$ shown in Fig.~\ref{fig3}{\bf a} is defined by $z = e^{iE}$, with $z$ being the eigenvalue of $W_{\lambda_1,\lambda_2,\eta}$. The winding number from Eq.~\eqref{eq:w} is non-trivial as there exists $z \in \partial \mathbb{D}$ such that $|\nu_\eta(z)|=1$. We confirm the breaking of PT symmetry by measuring the time evolution of the overall probability of photons, which is defined as $P(t)=\sum_x |\langle x|\psi(t)\rangle|^2$. In the PT-broken regime, one expects $P(t)$ to increase exponentially over time due to the positive imaginary components of the quasienergies. This behavior is experimentally confirmed in Fig.~\ref{fig3}{\bf b}, where the time evolution for arbitrary initial states indicates PT symmetry breaking. Moreover, we observe clear directional dynamics, as shown in Fig.~\ref{fig3}{\bf c}, where the walker's wave packet moves persistently to the right. The experimental results show a similarity $\mathcal{S} > 0.948$ compared with numerical simulations. Such directional flow can be further analyzed through the properties of the Lyapunov exponent $L_{\lambda_1,\lambda_2,\eta}$~\cite{WZ1,WZ2,qhlee} (see Methods).

Conversely, in the insulating phase ($\lambda_{1}<\lambda_{2}$), PT symmetry is preserved at finite $\eta$ with the winding number $\nu_\eta(z)=0$, as shown in Fig.~\ref{fig3}{\bf d}. In this regime, the measured $P(t)$ remains approximately constant, indicating that the quasienergies are entirely real. This observation aligns well with numerical simulations, as shown by the symbols and curves in Figs.~\ref{fig3}{\bf b}, {\bf e}. Furthermore, the dynamics remain localized, as depicted in Fig.~\ref{fig3}{\bf e}, in contrast with the dynamics in metallic phase. The measured similarity $\mathcal{S}>0.968$ in Fig.~\ref{fig3}{\bf f} supports our findings. The above discussion corroborates our complete experimental portrayal of the PT symmetry transition and localization transition in regions with small $\eta$.

\begin{figure}[h]
\includegraphics[width=0.5\textwidth]{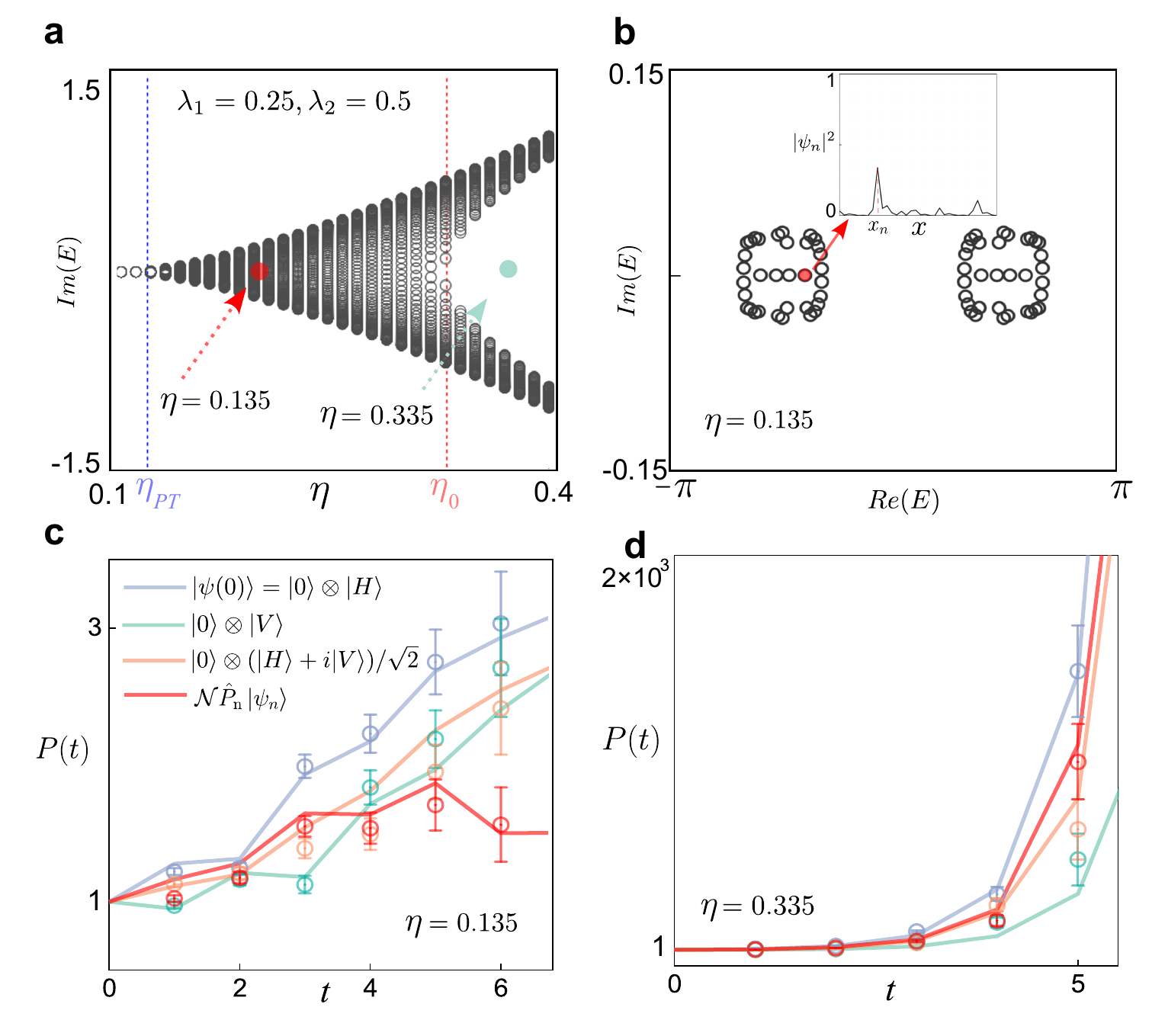}
\caption{{\bf Observation of the second phase transition in the non-Hermitian AAH model.} {\bf a} Imaginary parts of the quasienergies of a non-Hermitian quantum walk with $\theta=0$ for $\lambda_1=0.25<\lambda_2=0.5$. {\bf b} Numerically calculated quasienergy spectra in the complex plane for $\eta = 0.135$ under periodic boundary conditions. The inset shows the spatial distribution of the no-loss eigenstates. The other parameters are the same as those in {\bf a}. {\bf c}, {\bf d}  Measured overall probabilities $P(t)$ of the quantum walk with different initial states for $\eta=0.135$ and $\eta=0.335$, respectively.}\label{fig4}
\end{figure}

\minisubsection{Observation of the novel spectral transition in the PUAMO model}

Furthermore, as we further increase the non-Hermitian parameter $\eta$, the second phase transition is observed, whereby the quasienergy becomes entirely complex, as demonstrated in Fig.~\ref{fig4}{\bf a} (refer also to the green dots in Fig.~\ref{fig1}{\bf a}). After the second transition, the winding number becomes $|\nu_\eta(z)|=1$ for all $z \in \partial \mathbb{D}$. By analyzing the transfer matrix of $W_{\lambda_1=0.25,\lambda_2=0.5,\eta}$, two critical points are identified, $\eta_{\text{PT}}=0.119$ and $\eta_0=0.328$, corresponding to the PT symmetry transition and the second phase transition, respectively (see Supplementary Material). These two transition points are marked with dashed lines in Fig.~\ref{fig4}{\bf a}. To experimentally observe the second phase transition, we first note that when $\eta_{\text{PT}}<\eta=0.135<\eta_0$, there exist some quasienergies of $W_{\lambda_1,\lambda_2,\eta}$ that remain purely real. The eigenstates corresponding to these quasienergies are localized in real space and are denoted as ``no-loss states'', as illustrated in Fig.~\ref{fig4}{\bf b}. We choose an initial state that has a significant overlap with one of the no-loss states $|\psi_n\rangle$, $|\psi(0)\rangle = \mathcal{N} \hat{P}_{\text{n}} |\psi_n\rangle$, where $\hat{P}_{\text{n}} = |x_{n}\rangle \langle x_{n}|$ is the projection operator, and $x_n$ is the position of the maximum occupation probability of $|\psi_n\rangle$. The factor $\mathcal{N}$ ensures the normalization of the initial state. Since the no-loss state is primarily localized at $x_n$, its overlap with the initial state is significantly larger than with any other bulk state. Consequently, after $t$ steps, the time evolution of the state is $|\psi(t)\rangle = W^t |\psi(0)\rangle = \sum_{m}  e^{-iE_mt}|\psi_m\rangle\langle \psi_m |\psi(0)\rangle \approx e^{-iE_{n}t}|\psi_n\rangle$ in the short-time dynamics, where $E_n$ represents the quasienergy of the no-loss state. This scenario ensures that the overall photon probability does not significantly increase over time, which can be experimentally verified in Fig.~\ref{fig4}{\bf c}. Here, the measured $P(t)$ (red line) does not increase with time, whereas for other choices of initial states, $P(t)$ consistently and rapidly increases.

However, when the non-Hermitian parameter exceeds the second critical value, with $\eta=0.335>\eta_0$, the above scenario is no longer valid because all quasienergies acquire imaginary components and the evolved states always exhibit strong amplification, as shown in Fig.~\ref{fig4}{\bf d}. As a result, the contrasting behaviors of the measured $P(t)$ for $\eta < \eta_0$ (Fig.~\ref{fig4}{\bf c}) and $\eta > \eta_0$ (Fig.~\ref{fig4}{\bf d}) confirm the second phase transition within our system.

\minisection{Conclusion}

We demonstrate the first experimental realization of the (P)UAMO, a phenomenological analog to the (non-)Hermitian AAH model, through discrete-time quantum walks of single photons.
In the Hermitian regime, we validate the localization-delocalization transitions inherent to the AAH model, clearly distinguishing metallic and insulating phases by dynamic signatures. By extending our exploration to non-Hermitian regimes, we experimentally confirm and characterize two distinct phase transitions: the PT symmetry-breaking transition and a previously unobserved spectral transition beyond which all quasienergies become complex. We elucidate these transitions through dynamic observables and directly associate their emergence with changes in the spectral winding number. Moreover, the versatility of our system suggests practical applications beyond fundamental research. For instance, exploiting the observed PT-symmetry breaking transitions, our setup can be effectively employed for mode-selective amplification in laser systems. In this context, amplification can be confined exclusively to a targeted mode, thus leaving the behavior of other modes unchanged. This approach has significant potential for designing advanced optical devices with enhanced mode discrimination capabilities, further underscoring the applicability of non-Hermitian quantum walks to technological innovations in photonics. Our quantum-walk platform not only provides a versatile framework for experimental studies of non-Hermitian phenomena but also opens pathways toward engineering synthetic topological materials with customized transport properties.

\minisection{Methods}
\minisubsection{Experimental implementation}

As illustrated in Fig.~\ref{fig1}{\bf b}, the walker photon is initialized in a localized spatial mode and projected onto a selected polarization state through a sequence of polarization optics comprising a PBS, a QWP, and an HWP. The coin operator $Q_x$ is implemented by a set of wave plates
\begin{equation}
Q_x = q\left(\phi_3\right) h\left(\phi_2\right) q\left(\phi_1\right),
\end{equation}
where $q(\cdot)$, $h(\cdot)$ denote the unitary transformations implemented by QWP and HWP, respectively, with
\begin{align}
h(\phi) &=
\left[
\begin{array}{cc}
\cos 2\phi & \sin 2\phi \\
\sin 2\phi & -\cos 2\phi
\end{array}
\right], \\
q(\phi) &=
\left[
\begin{array}{cc}
\cos^2 \phi + i \sin^2 \phi & (1 - i)\sin\phi\cos\phi \\
(1 - i)\sin\phi\cos\phi & \sin^2 \phi + i \cos^2 \phi
\end{array}
\right].
\end{align}
The corresponding angles are $\phi_1=\frac{\pi}{4}-(x \Phi+\theta) \pi$, $\phi_2=\frac{\pi}{4}-\frac{1}{2} \arccos \lambda_2$, $\phi_3=\frac{\pi}{4}+(x \Phi+\theta) \pi$.

The shift operator $S_{\lambda_{1},\eta}$ is implemented by BDs, HWPs, and PPBSs, which is
\begin{equation}
S_{\lambda_{1},\eta}=e^{4\pi\eta}M_{E}S'_{2}h(\theta_2)S'_{1}h(\theta_1)M_{E}
\end{equation}
with $\theta_1=0$, $\theta_2=\arccos \lambda_1$. The conditional shift operators $S'_1$ and $S'_2$ move the walker in the corresponding coin states $|V\rangle$ and $|H\rangle$ to the right and left, respectively, with
\begin{align*}
S'_1 &= \sum_x |x+1\rangle\langle x| \otimes |V\rangle\langle V| + |x\rangle\langle x| \otimes |H\rangle\langle H|, \\
S'_2 &= \sum_x |x-1\rangle\langle x| \otimes |H\rangle\langle H| + |x\rangle\langle x| \otimes |V\rangle\langle V|.
\end{align*}
The mode-selective loss operator is given by  $M_E = \sum_{x}|x\rangle\langle x|\otimes\left[ \begin{array}{cc} 1 & 0 \\ 0 & e^{-2\pi\eta} \end{array} \right]$, and is experimentally implemented using a PPBS. After the application of $M_E$ at each step, photons in the vertical polarization state $|V\rangle$ are reflected by the PPBS with probability $p = 1-e^{-8\pi\eta}$, while the remaining photons continue to propagate in the quantum-walk evolution. The experimentally realized time-evolution operator ${\tilde W}$ differs from $W$ by a factor of $e^{4\pi \eta}$, i.e., $W=e^{4\pi \eta} {\tilde W}$~\cite{xiao1}.

In the detection stage, the probability distribution of the walker is obtained as $P(x,t) = \frac{N(x,t)}{\sum_x N(x,t)}$, where $N(x,t) = N_H(x,t) + N_V(x,t)$ is the total number of photons detected at position $x$ after the $t$th step. Here, $N_{H}(x,t)$ and $N_{V}(x,t)$ denote the photon counts with horizontal and vertical polarizations, respectively, measured by APDs. The overall probability $P(t)$ is reconstructed as
\begin{equation}
P(t) = e^{8\pi\eta t} \sum_x \frac{N(t, x)}{\sum_x \left[ N(t, x) + \sum_{t'=1}^{t} N_L(t', x) \right]},
\end{equation}
where $N_L(t, x)$ denotes the photon loss at position $x$ due to the partial measurement $M_E$ at time step $t$.
\smallskip

\minisubsection{PT-symmetry of the PUAMO}

PT-symmetry plays a significant role in characterizing phase transitions of the PUAMO: The Floquet operator $W_{\lambda_1,\lambda_2,\eta}$ in the modified ``time frame'' $\tilde{W}_{\lambda_1,\lambda_2,\eta}=Q_{\lambda_2,\theta}^{1/2}S_{\lambda_1}Q_{\lambda_2,\theta}^{1/2}$ exhibits PT-symmetry \cite{cedzi2}, that is, $(\mathcal{PT})\tilde{W}_{\lambda_1,\lambda_2,\theta,\eta}(\mathcal{PT})^{-1}=\tilde{W}_{\lambda_1,\lambda_2,\theta,\eta}^{-1}$. Here, $\mathcal{PT}$ represents the combined inversion of space and time given as the antilinear operator
\begin{equation*}
    \mathcal{PT}=\sum_x\ketbra x{-x}\otimes \mathcal{PT}_x\mathcal K,
\end{equation*}
acting locally as $\mathcal{PT}_x = \sigma_{z}$ where $\mathcal K$ is complex conjugation. As the non-Hermitian parameter $\eta$ increases, this PT-symmetry is spontaneously broken at the value $\eta=-\log\lambda_0$, where parts of the spectrum of $W_{\lambda_1,\lambda_2,\eta}$ move off of the unit circle, as illustrated in Fig.~\ref{fig1}{\bf a}.

Changing the time frame from $W_{\lambda_1,\lambda_2,\eta}$ to $\tilde W_{\lambda_1,\lambda_2,\eta}$ does not change the dynamics up to an initial rotation of the initial state. Also note that the original time frame is PT-symmetric, albeit for a position-dependent symmetry operator.
\smallskip

\minisubsection{Non-Hermitian localization-delocalization transition}

The significance of the Lyapunov exponent $L$ is that it quantifies the inverse localization length of solutions to the secular equation, that is, it determines whether or not eigenstates are localized ($L>0$) or not ($L=0$). With this in mind, the delocalization in $\eta$ can be understood: First, note that $W_{\lambda_1,\lambda_2,\eta}$ is similar to $W_{\lambda_1,\lambda_2,0}$ via the (non-unitary) skin transformation $V_\eta=\sum_x\sum_{s=0,1}e^{2\pi\eta x}\ketbra xx\otimes\ketbra ss$. Thus, if $\psi$ is an eigenstate of $W_{\lambda_1,\lambda_2,0}$, $(V_\eta^{-1}\psi)_x=e^{-2\pi\eta x}\psi_x$ is an eigenstate of $W_{\lambda_1,\lambda_2,\eta}$.

In the regime $\lambda_1<\lambda_2$, every eigenstate of $W_{\lambda_1,\lambda_2,0}$ is localized around some localization center $x_0$ and decays exponentially with inverse localization length $L=L_{\lambda_1,\lambda_2}=\log\lambda_0$, that is, $\psi_x\sim e^{-L|x-x_0|}$. Increasing $\eta$ leads to eigenstates $\psi^{(\eta)}$ of $W_{\lambda_1,\lambda_2,\eta}$ which by the above discussion follow
\begin{equation*}
    \psi^{(\eta)}_x\sim\begin{cases}
        e^{-(L+2\pi\eta)(x-x_0)}, & x>x_0,\\
        e^{(L-2\pi\eta)(x-x_0)}, & x<x_0.
    \end{cases}
\end{equation*}
For $\eta>0$, such eigenfunctions decay to both sides of $x_0$ as long as $2\pi\eta<L$. However, as soon as $2\pi\eta>0$, delocalization sets in to the left of $x_0$, while the eigenfunction remains localized to the right. This is precisely what we observe in the experiment, see Fig.~\ref{fig3}{\bf c}.

\smallskip

{\bf Data availability}

The data that support the findings of this study are available from the corresponding authors upon requests.

{\bf Code availability}

The codes that support the findings of this study are available from the corresponding authors upon requests.

{\bf Acknowledgments}

Q.L. and P.X. are supported by the National Key R$\&$D Program of China (Grant No. 2023YFA1406701) and the National Natural Science Foundation of China (Grants No. 12025401, No. 92265209). Q.Z. is supported by National Key R\&D Program of China (2020 YFA0713300) and Nankai Zhide Foundation.

{\bf Author contributions}

Q.L. performed the experiments and wrote part of the paper. C.C. and Q.Z. developed the theoretical aspects and wrote part of the paper. P.X. supervised the project, designed the experiments, analyzed the results, and revised the paper.

{\bf Competing interest declaration}

The authors declare no competing interests.

\ifarxiv
\begin{widetext}
\appendix
\section*{Appendix}
\subsection{Quantifying the phase transitions in the non-Hermitian regime.}

Central to understanding and quantifying the two phase transitions in the non-Hermitian regime is the understanding of the behavior of the Lyapunov exponent $L_{\lambda_1,\lambda_2,\eta}^\sharp$ of the dual PUAMO $W_{\lambda_1,\lambda_2,\eta}^\sharp$ as a function of $\eta$. This dual model has the same eigenvalue spectrum as $W_{\lambda_1,\lambda_2,\eta}$, so if PT-symmetry spontaneously breaks for the dual model, it also breaks for the original model. In the dual model, $\eta$ plays the role of a complexified phase $\theta\mapsto\theta-i\eta$. This (complexified) Lyapunov exponent $L_{\lambda_1,\lambda_2,\eta}^\sharp$ is given by \cite{cedzi2}
\begin{equation*}
    L_{\lambda_1,\lambda_2,\eta}^\sharp=\max\Big\{0,-\log\lambda_0+2\pi|\eta|-2\pi\max\{0,|\eta|-\eta_0\}\Big\},
\end{equation*}
where $L_{\lambda_1,\lambda_2,0}^\sharp=\max\{0,-\log\lambda_0\}$ with $\lambda_0$ as in Eq.~(3).
Here, $\eta_0=\arcsinh(\lambda_1'/\lambda_1)/(2\pi)$ quantifies the value of $\eta$ for which the associated transfer matrices are not analytic, a phenomenon that does not occur in the non-Hermitian AAH model or its dual. Since the derivation of this requires a lot of theoretical background, we refer the reader to \cite{cedzi,cedzi2} instead of reproducing it here.

In the extended and critical phase with $\lambda_1>\lambda_2$ and $\lambda_1=\lambda_2$, respectively, the Lyapunov exponent of the UAMO vanishes, that is, $L_{\lambda_1,\lambda_2,0}=\max\{0,\log\lambda_0\}=0$. The graph of $L_{\lambda_1,\lambda_2,\eta}^\sharp$ therefore has one turning point at $\eta=0$ and $\eta=\eta_0$, see Fig.~\ref{fig:lyap_exp}. Therefore, PT-symmetry is immediately spontaneously broken for every positive $\eta$.

In the localized phase $\lambda_1<\lambda_2$ the graph of $L_{\lambda_1,\lambda_2,\eta}^\sharp$ has four turning points at $\eta=\pm\lambda_0\neq0$ and $\eta=\pm\eta_0$
, see Fig.~ \ref{fig:lyap_exp}. One has $L_{\lambda_1,\lambda_2,\eta}^\sharp=0$ as long as $|\eta|<\log\lambda_0$. In this regime, the dual UAMO is delocalized, so the original model is localized. Moreover, in this regime, not only is the spectrum confined to the unit circle by PT-symmetry, it is also independent of $\eta$~\cite{cedzi}. At $\eta=\log\lambda_0$, the critical value of the first phase transition, PT-symmetry spontaneously breaks. As $|\eta|$ is increased above this value, the dual UAMO localizes, which implies delocalization for the original model, see Methods. The second critical value is given by $\eta=\eta_0$. It signals the value of $\eta$ beyond which $L_{\lambda_1,\lambda_2,\eta}^\sharp=-\log\lambda_0+\eta_0=\log[(1+\lambda_2')/\lambda_2]$ is independent of $\eta$ and $\lambda_1$, and only depends on $\lambda_2$. Physically, this corresponds to the value of $\eta$ for which either the (11)- or the (22)-entry of the coin of the dual model $W_{\lambda_1,\lambda_2,\eta}^\sharp$ vanishes infinitely often and therefore the walk decouples either to the left or to the right.

\begin{figure}[h]
    \includegraphics[width=.9\textwidth]{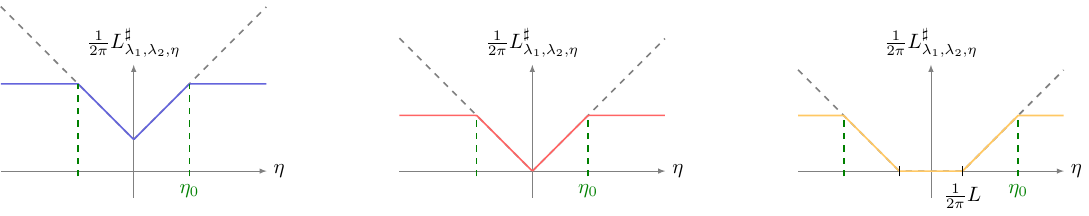}
    \caption{\label{fig:lyap_exp}The graph of the dual Lyapunov exponent $L^\sharp$ on the spectrum as a function of $\eta$. In the left panel, $\lambda_1<\lambda_2$ and there are two turning points: one at the Lyapunov exponent $L$ and one at $\eta_0=\arcsinh(\lambda_1'/\lambda_1)/(2\pi)$. These correspond to the critical values of the first and second phase transitions of the PUAMO, respectively. In the middle and right panels, $\lambda_1=\lambda_2$ and $\lambda_1>\lambda_2$, respectively, and there is only one transition at $\eta_0$. In particular, there is no localization-delocalization transition. The dashed lines show the graphs of the dual Lyapunov exponents for the corresponding non-Hermitian AAH model.}
\end{figure}

\subsection{Phenomenological correspondence between the (P)UAMO and the (non-Hermitian) AAH model}

In a certain regime of the non-Hermitian parameter $\eta$, the PUAMO model from Eq.~(1) is phenomenologically equivalent to the non-Hermitian AAH model given by
\begin{equation*}
    (H_{\lambda,\eta}\psi)_n=e^{2\pi \eta}\psi_{n+1}+e^{-2\pi \eta}\psi_{n-1}+2\lambda\cos(2\pi (\Phi n+\theta))\psi_n,
\end{equation*}
which is also known as the quasiperiodic Hatano-Nelson model~\cite{HatanoNelson1,HatanoNelson2,yliu}.
On the spectrum, the Lyapunov exponent of this model is given by \cite{bourgainContinuityLyapunovExponent2002,yliu}
\begin{equation*}
    L_{\lambda,\eta}=\log\lambda+2\pi|\eta|.
\end{equation*}
The graph of the dual Lyapunov exponent $L_{\lambda,\eta}^\sharp=L_{-\lambda,\eta}$ is shown in Fig.~\ref{fig:lyap_exp} as gray dashed lines.
In the Hermitian setting with $\eta=0$, this model has a metal-insulator transition at $\lambda=1$~\cite{Jitomirskaya1999Annals}, which in the non-Hermitian setting is shifted to $\eta=-\log\lambda$ \cite{yliu}.

The regime in which the PUAMO behaves analogously to this model is given by $|\eta|<\arcsinh(\lambda_1'/\lambda_1)/(2\pi)$. For larger values of $\eta$, eigenfunctions of the non-Hermitian AAH model with corresponding real eigenvalues continue to exist, whereas in this regime the PUAMO does not have eigenstates corresponding to eigenvalues on the unit circle anymore, compare the two panels in Fig.~\ref{fig:spectra_NHAAH_PUAMO}.

\begin{figure}[t]
    \begin{center}
        \includegraphics[width=0.95\textwidth]{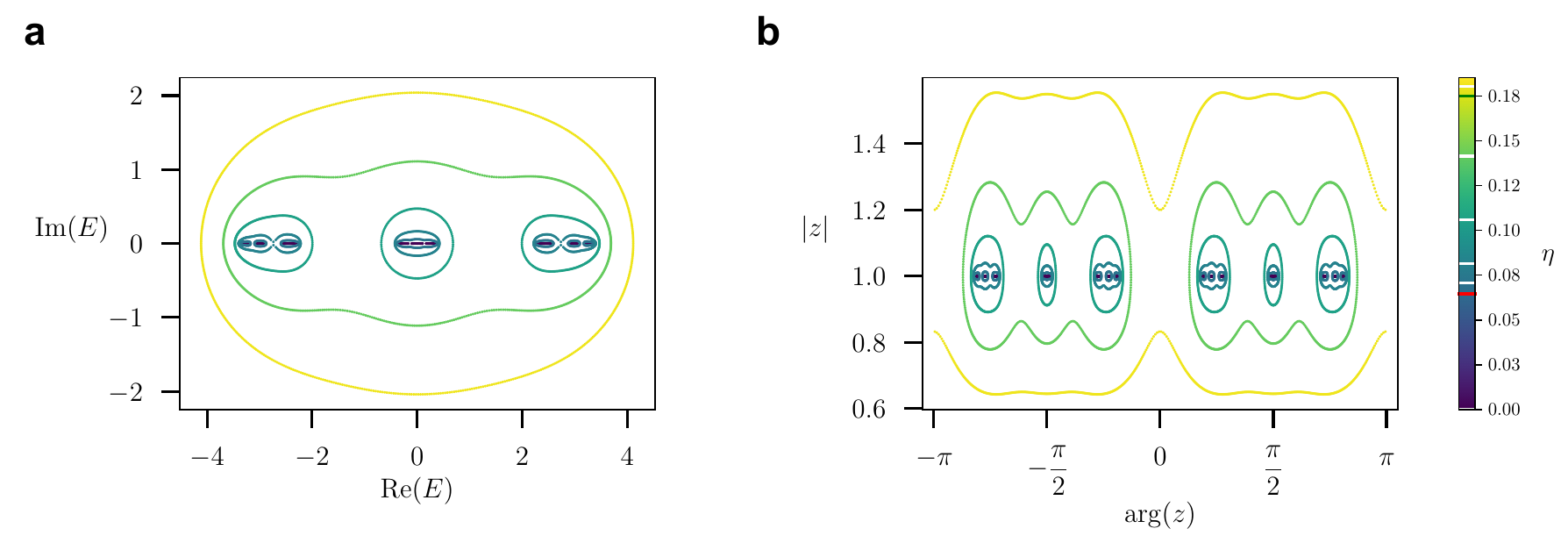}
    \end{center}
    \caption{\label{fig:spectra_NHAAH_PUAMO} {\bf a} The spectra of the non-Hermitian AAH model, and {\bf b} the PUAMO for varying $\eta$ where $E$ and $z$ correspond to the energy and the quasienergy, respectively. Here, the coupling constants $\lambda$ for the non-Hermitian AAH model and $\lambda_1$, $\lambda_2$ for the PUAMO are chosen such that $\lambda=1.5=\lambda_0$ with $\lambda_0$ as in Eq.~(3). The white markers in the colorbar indicate the displayed values of $\eta$ in both plots, and the red and green markers indicate the critical values of the first and second phase transition, respectively. Clearly, for $\eta$ larger than the first critical value, parts of the spectrum move off of the real axis and the unit circle, respectively. In the right plot, for $\eta$ larger than the value of the second phase transition, no spectrum is left on the unit circle. This phase transition cannot happen in the non-Hermitian AAH model, as it would correspond to a discontinuity of the spectrum with respect to $\eta$.}
\end{figure}

\subsection{Observation of the localization-delocalization transition in the PUAMO model.}

In the main text, we discuss the localization-delocalization transition induced by non-Hermiticity, which can be quantified by the Lyapunov exponent $L_{\lambda_1,\lambda_2,\eta}$. However, instead of fixing $\lambda_1$ and $\lambda_2$ and varying $\eta$, we can also observe the (same) phase transition for a fixed value of $\eta$ by tuning $\lambda_1$ and $\lambda_2$. The rationale behind this is that in the parameter region of localization, the Lyapunov exponent must be positive, that is,
\begin{equation*}
    L_{\lambda_1,\lambda_2,\eta}=\max\{0,\log\lambda_0-2\pi|\eta|\}>0,
\end{equation*}
with $\lambda_0$ as in Eq.~(3).
For fixed $\eta>0$, this is to require that $\log\lambda_0>2\pi|\eta|$, or, differently put, that
\begin{equation*}
    \lambda_2>\frac{2e^{2\pi\eta}\lambda_1(1+\lambda_1')}{2(1+\lambda_1')+(e^{4\pi\eta}-1)\lambda_1^2}.
\end{equation*}
The phase transition is thus expected at
\begin{equation}\label{eq:nh-phase_transition}
    \lambda_2=\begin{cases}\frac{2e^{2\pi\eta}\lambda_1(1+\lambda_1')}{2(1+\lambda_1')+\lambda_1^2(e^{4\pi\eta}-1)}, & 0\leq\lambda_1\leq\frac{2e^{2\pi|\eta|}}{1+e^{4\pi|\eta|}},\\
    1,&\frac{2e^{2\pi|\eta|}}{1+e^{4\pi|\eta|}}<\lambda_1\leq1.
    \end{cases}
\end{equation}
Note that for the UAMO with $\eta=0$, this boils down to the known phase transition at $\lambda_1=\lambda_2$~\cite{cedzi}.

To gain further insight into this modified phase transition of the PUAMO, we experimentally measure the second moment $\langle x^2 \rangle = \sum_x x^2 P(x, t)$ as a function of the parameters $\lambda_1$ and $\lambda_2$ across the entire parameter space. The second moment quantifies the spreading rate of the walker, thus providing a direct characterization of the system's localization properties. As shown in Fig.~\ref{fig:s3}{\bf a}, our experimental results clearly capture the localization transition, and these observations agree well with the theoretical result presented in Fig.~\ref{fig:s3}{\bf b}. Comparing Fig.~2{\bf h} and Fig.~\ref{fig:s3}{\bf a}, we observe that the introduction of non-Hermiticity modifies the boundary of the localization transition, which is consistent with the theoretical prediction given by Eq.~\eqref{eq:nh-phase_transition}.

\begin{figure}[h]
    \includegraphics[width=.7\textwidth]{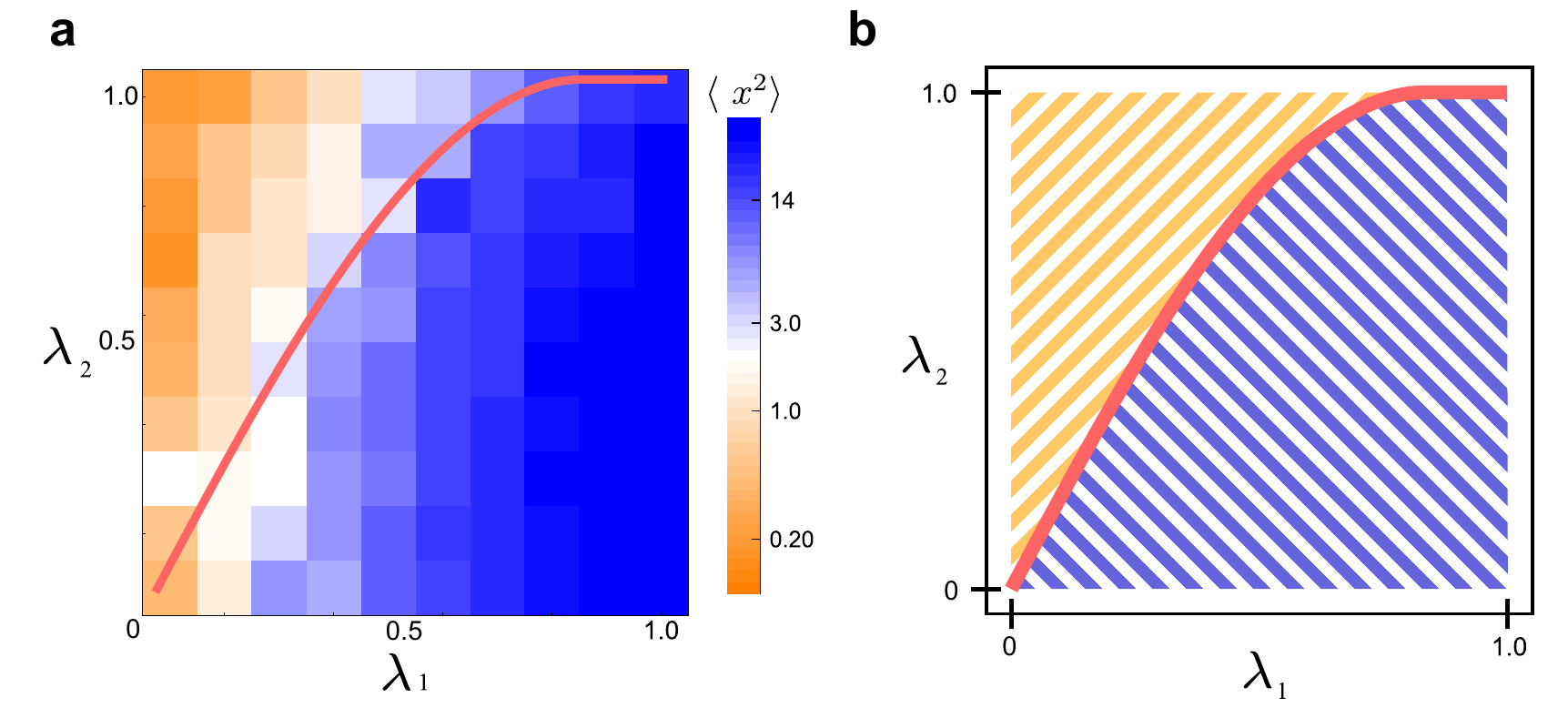}
    \caption{\label{fig:s3} {\bf a} Measured phase diagram, which is evaluated via the second moment of position $\langle x^2 \rangle$ after $t=6$ time steps. We set $\theta=0$, $\eta=0.1$ and chose $\ket{\psi(0)}=\ket{0}\otimes(\ket{H}+i\ket{V})/\sqrt{2}$ as the initial state. The dashed curve indicates the critical value of the first phase transition. {\bf b} Theoretical phase diagram with the phase transition in red according to Eq.~\eqref{eq:nh-phase_transition}.}
\end{figure}
\end{widetext}
\fi

\end{document}